\documentclass[sigconf,screen]{./required/acmart}
\usepackage{graphicx} 
\usepackage{microtype}
\title{Building Software Engineering Capacity through a University Open Source Program Office}

\setcopyright{rightsretained}
\acmDOI{10.1145/3663529.3663866}
\acmYear{2024}
\copyrightyear{2024}
\acmSubmissionID{fsecomp24poster-p2-p}
\acmISBN{979-8-4007-0658-5/24/07}
\acmConference[FSE Companion '24]{Companion Proceedings of the 32nd ACM International Conference on the Foundations of Software Engineering}{July 15--19, 2024}{Porto de Galinhas, Brazil}
\acmBooktitle{Companion Proceedings of the 32nd ACM International Conference on the Foundations of Software Engineering (FSE Companion '24), July 15--19, 2024, Porto de Galinhas, Brazil}

\author{Ekaterina Holdener}
\affiliation{ 
    \institution{Saint Louis University}
    \city{Saint Louis}
    \state{MO}
    \country{USA}
}
\email{kate.holdener@slu.edu}
\author{Daniel Shown}
\affiliation{ 
    \institution{Saint Louis University}
    \city{Saint Louis}
    \state{MO}
    \country{USA}
}
\email{daniel.shown@slu.edu}
\date{May 2024}

\begin{document}

\begin{abstract}
This work introduces an innovative program for training the next generation of software engineers within university settings, addressing the limitations of traditional software engineering courses. Initial program costs were significant, totaling \$551,420 in direct expenditures to pay for program staff salaries and benefits over two years. We present a strategy for reducing overall costs and establishing sustainable funding sources to perpetuate the program, which has yielded educational, research, professional, and societal benefits.
\end{abstract}

\maketitle

\section{Introduction}
University software engineering education traditionally relies on classroom instruction and projects. However, these methods often lack the depth of real-world experience necessary for students' careers. Typical classroom projects focus on ``green field'' development, failing to simulate the challenges graduates will face professionally. Additionally, time constraints and limited project options hinder instructors' ability to adequately evaluate student performance, particularly in larger classes. Consequently, such projects often lack longevity and practical application beyond the semester.

To address these limitations, some educators are utilizing existing open source projects to offer students a more authentic software development experience. Yet, the quality of this experience depends on the engagement of project maintainers, posing scalability and consistency challenges.

Saint Louis University's (SLU) Computer Science department has pioneered a solution to these issues through the establishment of Open Source with SLU (OSS)~\cite{oss-main}, funded by the Alfred P. Sloan Foundation in 2022. OSS operates as a university-based Open Source Program Office (OSPO). OSS provides similar services to other university OSPOs, and also directly maintains a portfolio of open source software. Over its initial two years, OSS has developed an effective approach to student-driven software development, addressing the challenges of traditional university-level software engineering education and offering broader benefits to university researchers and the local community. The key differences between SLU's solution and other programs that emphasize real-world software engineering experiences and open source development~\cite{lean, oss-columbia, oss-nd} are: 1) SLU's focus on creating sustainable value for SLU researchers, 2) the near-peer mentorship structure, involving capstone instructor, program director, graduate students and undergraduates, 3) program's plan to achieve financial sustainability.

\section {Program Structure}
The program consists of three core groups: clients, program staff, and student developers. Clients of the program are predominantly SLU faculty who need custom software for their research. However, OSS also works with external clients, such as non-profit organizations, and is expanding the program to working with corporate sponsors. Clients submit their software requests, which are evaluated by the program staff using the evaluation criteria published on the OSS website for transparency. Selected projects are assigned a dedicated team of student developers. Team formation aligns with the start of the semester, drawing students from the Computer Science Capstone course — a requirement for all undergraduate Computer Science majors.

Each capstone team collaborates with an assigned tech lead, a Computer Science master's student employed by OSS. Together, they are responsible for project success. The tech lead defines the project scope, milestones, and goals, which must be achievable in about 8-12 hours of work per student during a two-week sprint. The tech lead creates GitHub issues in the project repository, provides technical oversight, mentorship, and team building.

The Computer Science Capstone is a two-semester sequence, earning students 2 course credits per semester. SLU students are expected to dedicate 2-3 hours of work per week outside of class time per credit hour. In a two-week sprint, each student resolves one GitHub issue using a trunk-based development process: self-assign an issue, create a branch, and submit a pull request. The tech lead reviews the pull request, provides feedback, and merges it when all concerns are addressed.

Individual accountability is achieved through grading: if a student successfully resolves their assigned sprint issue on-time, they get 1 point added toward their grade. The instructor only needs to see if each student's pull request got merged, while tech leads are responsible for the quality of the merged pull requests, based on whether the issue is fully addressed, code quality, and coding standards.

Students who consistently engage with the project and follow the process succeed in getting their pull requests merged on-time. However, some issues may require more work than anticipated, and students may be unable to complete their issue despite consistent engagement. In such cases, students receive full credit for the sprint and continue working on the same issue in the following sprint.

The program staff includes the OSS Program Director and the capstone instructor, in addition to the tech leads. The Program Director manages tech lead training, while the capstone instructor oversees student project responsibilities. Both roles ensure program continuity. To ensure project sustainability, teams adhere to open source standards, including comprehensive developer guides and well-documented GitHub issues. This ensures projects can accept contributions from the worldwide open source community. Projects transition from one team to another between academic years, providing continuity and realistic experiences as new teams build upon previously developed code bases, rather than starting from scratch each time.

\section {Program Cost}
The primary program costs stem from supporting the program director's salary and tech lead stipends. 
Tech leads receive a monthly \$2,500 stipend, reflecting an expected 20-hour workweek. In the first two years of operation, direct program costs to support staff salaries and fringes are \$551,420, averaging \$275,710 in annual direct costs to support the staff. These costs are expected to increase due to the need to increase staff salaries, and growth in the capstone course enrollment, which requires more projects and thus more tech leads. The department's freshmen enrollment surge a few years ago led to a significant increase in the Fall 2024 capstone enrollment, with 56 students already registered, compared to 36 in Fall 2023 and 31 in Fall 2022. Following the freshmen enrollment spike, the department has seen a steady rise in incoming students, indicating an ongoing increase in capstone enrollment in the future.

Sustaining this program involves expenses, and OSS is exploring avenues for financial viability. This includes integrating funded software development into research grant proposals, partnering with industry for sponsored projects, and optimizing costs by incorporating tech lead experiences into an elective course, satisfying graduate program degree requirements, instead of paying tech leads stipends. Educating faculty about software's research-enhancing potential will be pivotal for future financial success. To attract industry-funded projects, OSS is developing a program that will allow companies to sponsor capstone teams working on open-source projects that align with company's interests, gaining insight into potential hires while mentoring student teams. As computer science department enrollment grows, the capstone course must accommodate more students. Typically, each funded tech lead manages two capstone projects, resulting in proportional program costs. To offset this, a new graduate course, ``Developing Open Source Software Products'', will be introduced, requiring instructor approval to ensure qualified student participation. This course will offer training for tech leads in technical and non-technical aspects of software engineering. Tech leads will be assessed based on the timeliness of their feedback to the rest of the team, peer evaluation, and ability to deliver value to the client.



\section{Program Impact}
Currently, the OSS project portfolio~\cite{oss-main} encompasses 15 projects, which will carry forward into the upcoming academic year with fresh student teams. These projects address genuine needs with real clients, providing students with practical software engineering scenarios, preparing them for their careers. The program's open source nature significantly amplifies its impact. Student developers fortify their resumes by actively contributing code, participating in discussions on GitHub issues, and submitting pull requests. This engagement enriches the students' GitHub profiles, which in turn enhances their employability.

Predominantly, the program serves researchers requiring data collection, analysis, or process automation tools. The open source nature of these tools extends their value beyond the initial development phase. Researchers from other institutions can replicate, validate, or extend the work, leading to new collaborations and innovation. For instance, the Speech Transcription and Tagging software aids SLU speech-language pathologists (SLPs) in diagnosing children's speech errors. This tool benefits SLPs worldwide, with its open source framework facilitating feature requests and community-driven development.

Beyond academia, OSS extends its impact to the local community. The Shelter Volunteer Scheduling application streamlines volunteer management at emergency homeless shelters during severe weather conditions. This crucial tool replaces cumbersome Google Docs sign-up processes, ensuring security, visibility into commitments, and traceability. Integrated into the larger Get Help platform, this application enables efficient community support systems.

Moreover, OSS creates global value for software developers. Pi4Micronaut, a Java library compatible with the Micronaut framework, is used for web application development on Raspberry Pi computers. This tool empowers programmers to create IoT applications without specialized hardware knowledge, driving innovation and accessibility in the software development landscape.

Our next steps are creating course materials for the new graduate course, establishing detailed rubrics for pull request reviewers, securing funded projects, and assessing the scalability of the growing capstone class.

\bibliography{references}

\end{document}